\documentclass[a4paper]{jpconf}
\usepackage{graphicx}
\usepackage{cite}
\usepackage{color}
\usepackage{amssymb}
\usepackage{amsmath}
\usepackage{amsfonts}
\usepackage{srcltx}
\begin{document}
\title{Non-equilibrium Dynamics, Thermalization and Entropy Production}
\author{Haye Hinrichsen$^1$, Christian Gogolin$^2$, and Peter Janotta$^1$}
\address{$^1$ Universit\"at W\"urzburg,
Fakult\"at f\"ur Physik und Astronomie
         97074 W\"urzburg, Germany}
\address{$^2$ Institute for Physics and Astronomy,
         Potsdam University, 14476 Potsdam, Germany}
\ead{hinrichsen@physik.uni-wuerzburg.de}

\address{STATPHYS-KOLKATA VII Conference Proceedings}

\begin{abstract}
This paper addresses fundamental aspects of statistical mechanics such as the motivation of a classical state space with spontaneous transitions, the meaning of non-equilibrium in the context of thermalization, and the justification of these concepts from the quantum-mechanical point of view. After an introductory part we focus on the problem of entropy production in non-equilibrium systems. In particular, the generally accepted formula for entropy production in the environment is analyzed from a critical perspective. It is shown that this formula is only valid in the limit of separated time scales of the system's and the environmental degrees of freedom. Finally, we present an alternative simple proof of the fluctuation theorem.
\end{abstract}


\def\tot{{\rm tot}} 	
\def\sys{{\rm sys}}	
\def\env{{\rm env}}	

\def\S{S}		
\def\St{\S_\tot}
\def\Ss{\S_\sys}
\def\Se{\S_\env}
\def\ASt{\bar\S_\tot}
\def\ASs{\bar\S_\sys}
\def\ASe{\bar\S_\env}

\def\O{\Omega}		
\def\Ot{\O_\tot}
\def\Os{\O_\sys}
\def\Oe{\O_\env}

\def\s{s}		
\def\sp{{s'}}
\def\c{c}
\def\cp{{c'}}

\def\P{P}		
\def\Pc{\P_\c}
\def\Pcp{\P_\cp}
\def\Ps{\P_\s}
\def\Psp{\P_\sp}
\def\d{{\rm d}}

\def\Jccp{J_{\c\cp}}	
\def\Jcpc{J_{\cp\c}}
\def\Jssp{J_{\s\sp}}
\def\Jsps{J_{\sp\s}}

\def\w{w}		
\def\wssp{\w_{\s\sp}}
\def\wsps{\w_{\sp\s}}
\def\wccp{\w_{\c\cp}}
\def\wcpc{\w_{\cp\c}}

\def\C{[X_\c]}
\def\Cp{[X_\cp]}
\def\Nc{N_\c}
\def\Ncp{N_\cp}
\def\extent{\xi_{\c\cp}}
\def\affin{A_{\c\cp}}

\def\headline#1{{\vspace{5mm}\noindent\begin{small}\textbf{#1}\end{small}}\\[2mm]}
\def\smallcaption#1{\caption{\footnotesize #1}}


\section{Introduction}

Classical non-relativistic statistical physics is based on a certain set of postulates. Starting point is a physical entity, called \textit{system}, which is characterized by a set $\Os$ of possible configurations. Although this configuration space could be continuous, it is useful to think of it as a countable set of discrete microstates $\s\in\Os$. Being classical means that the actual configuration of the system is a matter of objective reality, i.e. at any time $t$ the system \textit{is} in a well-defined configuration $s(t)$. The system is assumed to evolve in time by spontaneous transitions $\s \to \sp$ which occur randomly with certain transition rates $\wssp>0$. This commonly accepted framework is believed to subsume the emerging classical behavior of a complex quantum system subjected to decoherence. 

Although in a classical system the trajectory of microscopic configurations is in principle measurable, an external observer is usually not able to access this information in detail. The observer would instead express his/her partial knowledge in terms of the probability $\Ps(t)$ to find the system at time $t$ in the configuration $\s$. In contrast to the unpredictable trajectory $\s(t)$, the probability distribution $\Ps(t)$ evolves deterministically according to the master equation $\dot\Ps(t) ~=~ \sum_{\sp\in\Os}(\Psp(t)\wsps-\Ps(t)\wssp)$.

\newpage
A system is said to \textit{equilibrate} if the probability distribution $\Ps(t)$ becomes stationary in the limit $t\to\infty$. In addition, if the probability currents $\Jssp=\Ps(t)\wssp$ and $\Jsps=\Psp(t)\wsps$ between all pairs of configurations $\s,\sp$ individually cancel each other in the stationary state, the system is said to \textit{thermalize}, obeying \textit{detailed balance}. The predictive power of equilibrium statistical mechanics relies on the fact that the stationary probability distribution of a thermalized system is universal and can be classified into a small number of thermodynamic ensembles. In particular, an isolated system thermalizes in such a way that each available configuration is visited with the same probability. This fundamental \textit{equal a priori probability postulate} is at the core of equilibrium statistical mechanics, from which all other thermodynamic ensembles can be derived.

\begin{figure}
\centering\includegraphics[width=50mm]{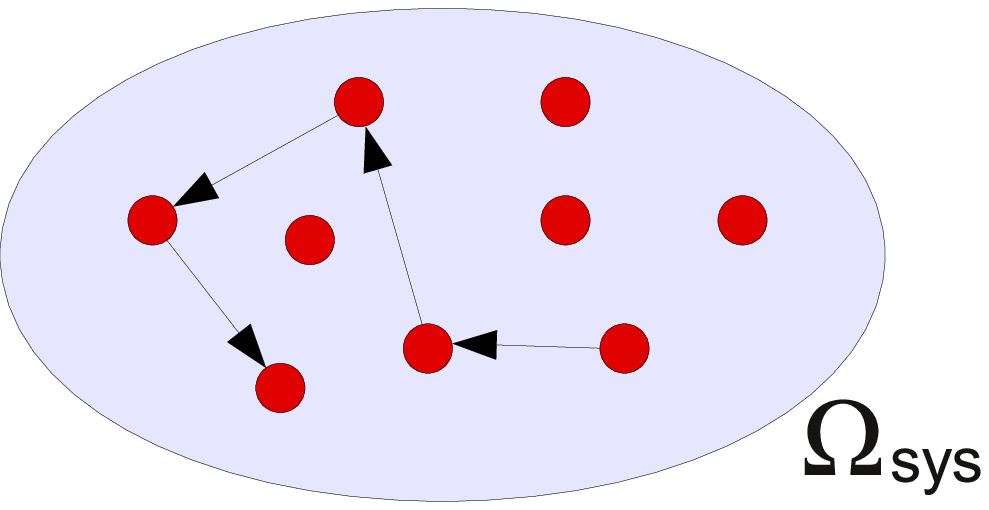}
\smallcaption{
Cartoon of a complex statistical system as a space $\Os$ of configurations (red dots). At any time the system \textit{is} in one of the configurations and evolves by spontaneous transitions (indicated by arrows) selected randomly according to specific transition rates.}
\end{figure}

Classical thermodynamics is concerned with time-dependent phenomena close to equilibrium. Introducing notions such as thermodynamic forces, potentials and currents it makes statements how such systems relax towards thermal equilibrium. Contrarily, non-equilibrium statistical physics deals with systems that do not thermalize, meaning that their probability currents do not vanish even in the stationary state. Typically such systems need to be driven from the outside in order to prevent them from thermalizing. 

The basic framework of statistical physics and thermodynamics sketched above is often taken for granted. However, one should be aware that the underlying postulates are highly non-trivial. In this paper we address some of these issues from a critical perspective. Where does the cartoon of spontaneous hopping between configurations come from? How can we be sure that the system will equilibrate into a stationary state? What is known about thermalization and the justification of the \textit{equal a priori probability postulate}? What is the meaning of non-equilibrium in an environment that thermalizes? What is behind the commonly accepted formula for the entropy production of non-equilibrium systems? In the following we make an attempt to discuss some of these questions in a common context, giving partial answers and pointing out open questions. Moreover, we would like to draw the readers attention to the fascinating links between classical statistical physics and recent developments in quantum information science.

The paper is organized as follows. In the following section we summarize existing textbook knowledge on stochastic system in a compressed form suitable for beginners in the field. Sect.~\ref{sec:quantum} deals with the question how we can justify the basic assumptions and postulates of classical statistical physics from the quantum perspective. In Sect. \ref{sec:entropy} we will discuss the problem of entropy production in non-equilibrium systems, analyzing the conditions under which commonly accepted formulas are valid. A discussion  of the fluctuation theorem together with an alternative compact proof is presented in Sect. \ref{sec:fluc}. The paper closes with concluding remarks in Sect.~\ref{sec:conclusions}.

\newpage
\section{Setup of classical statistical physics}
\label{sec:setup}

\headline{Configuration space and dynamics}
As outlined in the introduction, classical statistical physics is mainly concerned with models of complex systems having the following properties:
\begin{enumerate}
\item The system is characterized by a certain set $\Os$ of configurations $\s\in\Os$, also called microstates. Usually the configuration space is implicitly specified by the definition of a model. For example, in a reaction-diffusion model this space is the set of all possible particle configurations while in a growth model the microstates are identified with the possible configurations of an interface. 
\item The states are classical, i.e., at any time the system \textit{is} in one particular configuration $\s(t)$.
\item The system evolves randomly by instantaneous transitions $\s \to \sp$ occurring spontaneously with certain transition rates $\wssp\geq 0$. In numerical simulations, this dynamics is approximated by random-sequential update algorithms.
\end{enumerate}
Starting with an initial configuration $\s_0$ the system evolves randomly through an unpredictable sequence of configurations $\s_0 \to \s_1 \to \s_2 \to \ldots$ by instantaneous transitions. These transition take place at certain points of time $t_1,t_2,\ldots $ which are distributed according to a Poisson distribution like shot noise. Such a sequence of transitions is called a {\em stochastic path}. 

Although the actual stochastic path of the system is unpredictable, the probability $\Ps(t)$ to find the system in configuration $\s$ at time $t$ evolves deterministically according to the master equation 
\begin{equation}
\frac{\d}{\d t} \Ps(t) ~=~ \sum_{\sp\in\O}\Bigl(\Jsps(t)-\Jssp(t)\Bigr)\,,
\end{equation}
where
\begin{equation}
\Jssp(t)=\Ps(t)\wssp
\end{equation}
is the probability current flowing from configuration $\s$ to configuration $\sp$. The system is said 
\begin{itemize}
 \item to be \textit{stationary} or \textit{equilibrated} if the probability distribution $\Ps(t)$ is time-independent, meaning that for a given configuration $\s$ all incoming and outgoing probability currents cancel. \vspace{1mm}
 \item to \textit{equilibrate} if the master equation evolves into a stationary distribution in the limit $t \to \infty$, denoted as $\Ps=\Ps(\infty)$. 
\end{itemize}
For simplicity we will assume that this stationary state is unique and independent of the initial state. Note that systems with a finite configuration space always relax into a stationary state while for systems with an infinite or continuous configuration space a stationary state may not exist. Moreover, we will assume that the dynamics of the system under consideration is ergodic, i.e., the network of transitions is connected so that each configuration can be reached.

\newpage
\headline{Detailed balance}
A stationary system is said to \textit{thermalize} if it evolves into a stationary state which obeys \textit{detailed balance}. This means that the probability currents between all pairs of configurations cancel, i.e.
\begin{equation}
\label{eq:db}
\Jssp=\Jsps\qquad \forall \s,\sp\,,
\end{equation}
or equivalently
\begin{equation}
\label{eq:db2}
\frac\wssp\wsps\;=\;\frac\Psp\Ps \qquad \forall \s,\sp\,,
\end{equation}
where $\Ps$ is the stationary probability distribution. It is worth being noted that detailed balance can be defined in an alternative way without knowing the stationary probability distribution. To see this let us consider a closed loop of three transitions $\s_1 \to \s_2 \to \s_3 \to \s_1$. For these transitions Eq.~(\ref{eq:db2}) provides a system of three equations. By multiplying all equations one can eliminate the probabilities $\P_{s_i}$, arriving at the condition $w_{\s_1\s_2}w_{\s_2\s_3}w_{\s_3\s_1}=w_{\s_1\s_3}w_{\s_3\s_2}w_{\s_2\s_1}$. A similar result is obtained for any closed loop of transitions, hence the condition of detailed balance can be recast as
\begin{equation}
\label{eq:dbnew}
\prod_i w_{\s_i\s_{i+1}} \;=\; \prod_i w_{\s_{i+1}\s_{i}} 
\end{equation}
for all closed loops in $\Os$. A system with this property is said to have a \textit{balanced dynamics}.

Note that in a system with balanced dynamics we may rescale a pair of opposite rates by $\wssp\to\Lambda\wssp$ and $\wsps\to\Lambda\wsps$ without breaking the detailed balance condition. This intervention changes the dynamics of the model and therewith its relaxation, but the stationary state (if existent and unique) remains the same. This statement holds even if pairs of opposite rates are set to zero as long as this manipulation does not break ergodicity.

\headline{Equal a priori postulate}
Classical statistical physics is based on the so-called \textit{equal a priori postulate}. This postulate states that an isolated system, which does not interact physically with the outside world, will thermalize into a stationary state where all accessible configurations occur with the same probability, i.e. 
\begin{equation}
\Ps=const=1/|\Os|\,. \qquad\qquad\forall\s\in\Os
\end{equation}
It is assumed that this state obeys detailed balance since otherwise one could use the non-vanishing probability currents to construct a \textit{perpetuum mobile}. The immediate consequence would be that all transitions are reversible and that opposite rates coincide, i.e. $\wssp=\wsps$.

\headline{Entropy}
Entropy is probably the most fundamental concept of statistical physics. From the information-theoretic point of view, the entropy of a system is defined as the amount of information (measured in bits) which is necessary to describe the configuration of the system. Since the description of a highly ordered configuration requires less information than a disordered one, entropy can be viewed as a measure of disorder.

The amount of information which is necessary to describe a configuration depends on the already existing partial knowledge of the observer at a given time. For example, deterministic systems with a given initial configuration have no entropy because the observer can compute the entire trajectory in advance, having complete knowledge of the configuration as a function of time even without measuring it. Contrarily, in stochastic systems the observer has only a partial knowledge about the system expressed in terms of the probability distribution $\Ps(t)$. In this situation the amount of information needed to specify a particular configuration $\s\in\Os$ is $-\log_2 \Ps(t)$ bits, meaning that rare configurations have more entropy than frequent ones. 

Different scientific communities define entropy with different prefactors. In information science one uses the logarithm to base 2 so that entropy is directly measured in bits. Mathematicians instead prefer a natural logarithm while physicists are accustomed to put an historically motivated prefactor $k_B$ in front, giving entropy the unit of an energy. In what follows we set $k_B=1$, defining the entropy of an \textit{individual} configuration $\s$ as
\begin{equation}
\Ss(t,\s) = - \ln \Ps(t)\,.
\end{equation}
Since this entropy depends on the actual configuration $\s$, it will fluctuate along the stochastic path. However, its expectation value, expressing the observer's average lack of information, evolves deterministically and is given by
\begin{equation}
\Ss(t) = \langle\Ss(t,\s)\rangle_s = - \sum_{\s\in\Os} \Ps(t) \ln \Ps(t)\,,
\end{equation}
where $\langle \ldots\rangle$ denotes the ensemble average over independent realizations of randomness. Apart from the prefactor, this is just the usual definition of Shannon's entropy~\cite{Jaynes57,Jaynes57_2}. 

Up to this point entropy is just an information-theoretic concept for the description of configurations. The point where entropy takes on a \textit{physical} meaning is the \textit{equal a priori postulate}, stating that an isolated system thermalizes in such a way that the entropy takes the largest possible value $\Ss=\ln|\Os|$. As is well-known, all other thermodynamic ensembles can be derived from this postulate.

The numerical determination of entropies is a nontrivial task because of the highly non-linear influence of the logarithm. To measure an entropy numerically, one first has to estimate the probabilities $\Ps(t)$. The resulting symmetrically distributed sampling errors in finite data sets are amplified by the logarithm, leading to a considerable systematic bias in entropy estimates. Various methods have been suggested to reduce this bias on the expense of the statistical error, see e.g.~\cite{Grassberger,Bonachela}.

\headline{Subsystems}
In most physical situations the system under consideration is not isolated, instead it interacts with the environment. In this case the usual approach of statistical physics is to consider the system combined with the environment as a composite system. This superordinate \textit{total system} is then assumed to be isolated, following the same rules as outlined above. To distinguish the total system from its parts, we will use the suffixes `tot' for the total system while `sys' and 'env' refer to the embedded subsystem and its environment, respectively.

The total system is characterized by a certain space $\Ot$ of classical configurations $\c\in\Ot$ (not to be confused with system configurations $\s\in\Os$). The number of these configurations may be enormous and they are usually not accessible in experiments, but in principle there should be a corresponding probability distribution $\Pc(t)$ evolving by a master equation
\begin{equation}
\label{eq:cmaster}
\frac{\d}{\d t} \Pc(t) ~=~ \sum_{\cp\in\Ot}\Bigl(\Jcpc(t)-\Jccp(t)\Bigr)\,,\qquad\qquad\Jccp(t)=\Pc(t)\wccp
\end{equation}
with certain time-independent transition rates $\wccp\geq0$. 

\begin{figure}
\centering\includegraphics[width=100mm]{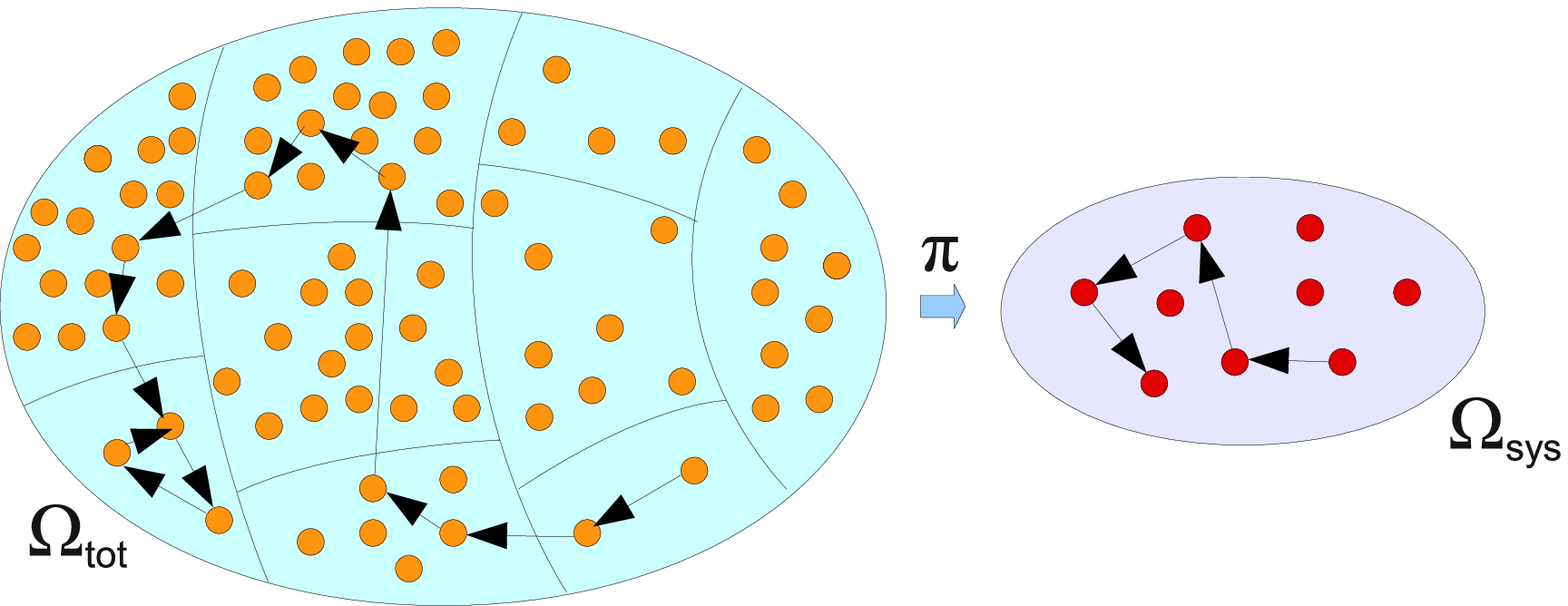}
\smallcaption{
A subsystem is defined by a projection $\pi$ which maps each state $\s \in \Ot$ of the total system (left) onto a particular state $\c\in\Os$ of the subsystem (right), dividing the state space of the total system into sectors. The figure shows a stochastic path in the total system together with the corresponding stochastic path in the subsystem.
}
\label{fig:sectors}
\end{figure}

Let us now consider an embedded subsystem. Obviously, for every classical configuration $\c\in\Ot$ of the total system we will find the subsystem in a well-defined unique configuration $\s\in\Os$. Conversely, for a given configuration of the subsystem $\s\in\Os$ the environment (and therewith the total system) can be in many different states. This relationship can be expressed in terms of a surjective map $\pi:\Ot\to\Os$ which projects every configuration $\c$ of the total system onto the corresponding configuration $\s$  of the subsystem, as sketched in schematically Fig.~\ref{fig:sectors}.

The projection $\pi$ divides the space $\Ot$ into sectors $\pi^{-1}(\c) \subset \Ot$ which consist of all configurations which are mapped onto the same $\s$. Therefore, the probability to find the subsystem in configuration $\s\in\Os$ is the sum over all probabilities in the corresponding sector, i.e.
\begin{equation}
\Ps(t) \;=\; \sum_{\c(\s)} \Pc(t)\,,
\end{equation}
where the sum runs over all configurations $\c\in\Ot$ with $\pi(\c)=\s$. Likewise, the projected probability current $\Jssp$ in the subsystem flowing from configuration $\s$ to configuration $\sp$ is the sum of all corresponding probability currents in the total system:
\begin{equation}
\Jssp(t) \;=\; \sum_{\c(\s)}\sum_{\cp(\sp)}  \Jccp(t) \;=\; \sum_{\c(\s)} \Pc(t) \sum_{\cp(\sp)}  \wccp\,.
\end{equation}
This allows us to define effective transition rates in the subsystem by
\begin{equation}
\label{eq:reducedrate}
\wssp(t) \;=\; \frac{\Jssp(t)}{\Ps(t)} \;=\; \frac{\sum_{\c(\s)} \Pc(t)\sum_{\cp(\sp)}\wccp}{\sum_{\c(\s)}\Pc(t)}\,.
\end{equation}
In contrast to the transition rates of the total system, which are usually assumed to be constant, the effective transition rates in the subsystem may depend on time. With these time-dependent rates the subsystem evolves according to the master equation
\begin{equation}
\label{eq:smaster}
\frac{\d}{\d t} \Ps(t) ~=~ \sum_{\sp\in\Os}\Bigl(\Jsps(t)-\Jssp(t)\Bigr)\, \qquad\qquad\Jssp(t)=\Ps(t)\wssp(t)\,.
\end{equation}
From the subsystems point of view this time dependence reflects the unknown dynamics in the environment. Moreover, ergodicity plays a subtle role: Even if the dynamics of the total system was ergodic, the dynamics \textit{within} the sectors $\pi^{-1}(\c)$ is generally non-ergodic and may decompose into several ergodic subsectors. As we will see in Sect.~\ref{sec:entropy}, this allows the environmental entropy to increase even if the subsystem is stationary.

\headline{Systems far from thermal equilibrium}
In Nature many systems are not thermalized but out of equilibrium. For this reason the study of non-equilibrium systems plays an increasingly important role in statistical physics. It should be noted that different communities use term `non-equilibrium' in a different way. In the context of thermodynamics, non-equilibrium refers to non-stationary situations close to thermal equilibrium, while statistical physicists use this term for systems violating detailed balance. In the following we use the nomenclature that a system is
\begin{itemize}
\item in \textit{thermal equilibrium} if its probability distribution is stationary obeying detailed balance.
\item \textit{thermalizing} if it relaxes towards thermal equilibrium with balanced rates obeying Eq.~(\ref{eq:dbnew}).
\item in a \textit{non-thermal equilibrium} if its probability distribution is in a stationary state without satisfying detailed balance. 
\item in \textit{out of thermal equilibrium} if it is non-stationary violating Eq.~(\ref{eq:dbnew}).
\end{itemize}

Since in Nature isolated systems are expected to thermalize, we can conclude that conversely a non-thermalizing system must always interact with the environment. This means that an external drive is needed to prevent the system from thermalizing, maintaining its non-vanishing probability currents. On the other hand, the total system composed of laboratory system and environment should thermalize. This raises the question how a thermalizing total system can contain a non-thermalizing subsystem?

The answer to this question is given in Eq.~(\ref{eq:reducedrate}). Even if the total system was predetermined to thermalize, meaning that the rates $\wccp$ obey Eq.~(\ref{eq:dbnew}), one can easily show that the effective rates $\wssp$ of transitions in the subsystem are generally not balanced. Therefore, a thermalizing `Universe' may in fact contain subsystem out of thermal equilibrium. The apparent contradiction is resolved by the observation that the projected rates $\wssp(t)$ depend on time: Although these rates may initially violate detailed balance, they will slowly change as the `Universe' continues to thermalize, eventually converging to values where they do obey detailed balance. This process reflects our everyday experience that any non-thermalizing system will eventually thermalize when the external drive runs out of power.

\section{Justification from the quantum perspective}
\label{sec:quantum}

The standard setup of statistical mechanics as described in section~\ref{sec:setup} is amazingly successful in explaining a wide range of physical processes. In stark contrast to this strong \emph{justification by corroboration} the question of whether and how it can be justified \emph{microscopically} is still open to a great extent. Over the course of the last century many famous physicists, such as Ludwig Boltzmann, have tried to derive statistical physics from Newtonian mechanics. Despite major efforts no fully convincing and commonly accepted microscopic foundation for statistical physics could be found \cite{Uffink}. Nowadays, quantum mechanics is the commonly accepted microscopic theory but a complete justification of statistical mechanics from quantum mechanics is yet to be achieved. This is surprising as already the founding fathers of quantum theory have worked on this problem~\cite{SchroedingerNeumann}.

Recently, stimulated by new experiments \cite{Bloch1,Bloch2} and novel methods from quantum information theory, a renewed interest in such fundamental questions ignited a flurry of activity in this field with significant new insights, leading to a reconsideration of the axiomatic system of statistical mechanics. As we shall see below, quantum mechanics and statistical physics apparently contradict each other but, very surprisingly, at the same time some genuine features of quantum mechanics can be used to justify essential assumptions of statistical mechanics. In this Section we discuss some of these recent developments, which we think are of interest to a wider audience, in a non-technical fashion. Readers not interested in the relationship of quantum mechanics and statistical physics can safely skip this section.

\headline{Emergence of the classical state space}
%
One aspect of the quantum mechanical foundations of statistical mechanics is to explain why macroscopic systems can be well described without considering the quantum mechanical nature at the microscopic level. In the following we discuss a general mechanism that explains how classical ensembles can emerge from the microscopic quantum dynamics.

As an example for an intrinsically quantum mechanical process that can be successfully described within the framework of statistical mechanics, let us consider the emission and absorption of light (photons) by atoms. This example includes both equilibrium situations, like the interaction of atoms with black body ration, and extreme nonequilibrium situation, like the pumping of a laser. The statement that emission and absorption of light can be described within the framework of statistical mechanics is to be understood in the following sense: quantum mechanics enters the description only in so far as it defines the energy levels of the atom. The relevant energy eigenstates of the quantum Hamiltonian are then taken to be the configurations $\s$ that constitute the classical state space $\Os$ of the atom while the dynamics is entirely described in terms of transition rates between these levels, corresponding to absorption, spontaneous emission, and stimulated emission. Although this description turned out to be useful, it is not at all clear why this simplified, classical treatment of-light matter interaction is eligible. As an essential feature, quantum mechanics allows systems to be in \emph{coherent superpositions} of energy eigenstates rather than \emph{probabilistic superpositions}. Why is it possible to simply ignore this fundamental feature of quantum mechanics and work with a classical description that only includes incoherent, \emph{probabilistic superpositions} of energy eigenstates and transitions between them?

The first step in resolving this problem is to realize that it is in general impossible to completely isolate a system from its environment. By tracing out the degrees of freedom in the environment it is easy to show that the time evolution of such an interacting system is not unitary anymore. In particular, the interaction with the environment can suppress coherent superpositions in the system, a process called \emph{decoherence} \cite{Zurek,Zeh96,Schlosshauer}. The insights obtained in the field of decoherence theory have proved to be extremely valuable, for both applications and from a fundamental perspective. However, most of the results are restricted to specific models or rely on assumptions such as a special form of the interaction or on approximations that are hard to control. Given the broad applicability of statistical mechanics and thermodynamics one would rather want to have an explanation for the emergence of classical state spaces from quantum theory that does not rely on such details.

Recently, building on earlier works \cite{Linden09,Linden10}, a result in this spirit was obtained in Ref.~\cite{PRE81}. It can be summarized informally as follows. Whenever a system interacts \textit{weakly} with an environment its quantum state (density operator) is close to being in a purely probabilistic superposition of energy eigenstates for most times. The result shows that quantum mechanics implies a generic mechanism leading to a suppression of coherent superpositions in such a way that an observer measuring the system at an arbitrary point in time is most likely to find the system in a state where only classical probabilistic superpositions contribute significantly. In other words, the coherent quantum dynamics of system plus environment suppresses coherence in the subsystem. The result can be expressed as a rigorous inequality bounding the off-diagonal entries in the density matrix of the subsystem. Its proof does not rely on any special properties of the interaction, it rather follows directly from the full unitary evolution of the total system without any approximations. The inequality is meaningful only if the interaction with the environment is \emph{weak}, meaning that it must be smaller (or at least not significantly larger) than the gaps between the energy levels of the system.

Can this result explain why the energy eigenbasis is the correct choice for the state space in the classical description of an atom absorbing and emitting light?
Thinking of the atom plus the surrounding light field as a quantum system subjected to decoherence by interaction with other atoms of the gas it does. In fact, the energy gaps of a few eV are several orders of magnitude larger than the thermal energy of the atoms (about 25 meV at room temperature), which sets the relevant energy scale for the coupling to other atoms of the gas. The weak perturbation caused by thermal scattering in the gas naturally leads to a decoherence into the energy eigenbasis, meaning that the classical description in terms of a probability $\Ps(t)$ to be in the energy eigenstate $\s$ at time $t$ describes the state of the system almost completely.

\begin{figure}
  \centering\includegraphics[width=110mm]{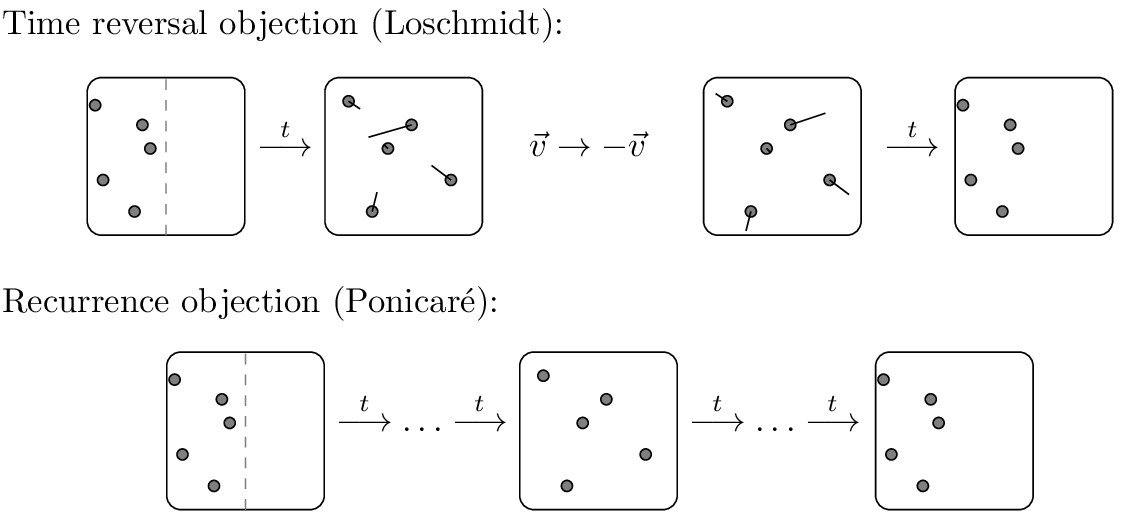}
  \smallcaption{
    Schematic depiction of the two objections against irreversibility in systems based on Newtonian mechanics. According to Loschmidt's \emph{time reversal objection}, for every process that increases the entropy of a system, say the spreading of particles in a container after removal of a wall, there exists a corresponding time-reversed process, which would reduce the entropy of the system. According to Ponicar\'{e}'s \emph{recurrence objection}, whenever the entropy of a system increased, the recurrent nature of Newtonian evolution implies that the entropy has to decrease after a sufficiently long period of time. The two objections illustrate that the Second Law in its usual form is incompatible with Newtonian mechanics. Equivalent objections apply to quantum mechanics which is also time reversal invariant and recurrent.}
  \label{fig:objections}
\end{figure}

\headline{Equilibration and apparent irreversibility}
\label{sec:equilibration}
Another important issue is an apparent incompatibility between quantum mechanics and statistical physics: While the Second Law, a corner stone of statistical physics, postulates the irreversible convergence of $\Ps(t)$ to a stable equilibrium distribution in the limit $t\to\infty$, the unitary dynamics of isolated quantum systems is \emph{time reversal invariant} and \emph{recurrent}; meaning that for every initial state the system will eventually return to an almost indistinguishable state after a certain (possibly extremely large) recurrence time. The famous objections brought forward by Loschmidt and Ponicar\'{e} against Boltzmann's H-Theorem, which was intended to be a proof of the Second Law for Newtonian mechanics, the so called \emph{time reversal objection} and the \emph{recurrence objection} (see Fig.~\ref{fig:objections}) equally apply if quantum mechanics is taken to be the underlying microscopic theory. 

These two objections show that true equilibration in the usual sense of statistical mechanics is impossible in quantum theory. The origin of the tendency to evolve towards equilibration, which constitutes a crucial part of the framework of statistical physics, on first sight, appears to be miraculous from the quantum perspective. However, recent studies show a way how it is possible to resolve this apparent contradiction. All we can hope for is to find equilibration in a weaker sense, meaning that the system is \emph{almost equilibrated for most times}. That is, the expectation values of a set of relevant observables could \emph{evolve towards} and then \emph{stay close} to a certain \emph{equilibrium value} for \emph{most times} during the evolution, even if the system started far from equilibrium. The notion of equilibration in classical statistical mechanics that comes closest to the quantum version of equilibrium seems to be what is called a \emph{non-thermal equilibrium} above.

Of particular interest are observables acting on a small part of a larger composite system. Surprisingly, as shown in a series of recent papers~\cite{Reimann08,Linden09,Short10}, it is possible to rigorously prove equilibration for almost all times for such observables from the unitary, time reversal invariant, and recurrent time evolution of quantum mechanics. More specifically, whenever the dimension of the Hilbert space of a subsystem is much smaller than a quantity called the effective dimension of the initial state, it is shown that the unitary dynamics of the full system is such that for most times the states of all small subsystems are practically indistinguishable from apparent equilibrium states. The proof requires that the Hamiltonian is fully interactive and non-degenerate, i.e., it does not decompose into non-interacting subsystems, -- a very weak and quite natural assumption.

The effective dimension, which is defined as $d_{\rm eff}(\rho) = 1/{\rm Tr} \rho^2$, measures how many eigenstates of the Hamiltonian contribute significantly to the initial state. Can we expect it to be large in realistic situations? In realistic many particle systems energy is approximately extensive. Even if we assume it to grow at most polynomially with the number of constituents, then an energy interval of fixed width will usually contain exponentially many energy eigenstates. This implies that even states with a very small energy uncertainty will usually consist of exponentially many energy eigenstates and thus it is safe to assume that the effective dimension will be very large in realistic situations. Thus we can conclude that equilibration of small subsystems in large, interacting quantum systems is a generic property.
Similar results can be obtained for certain sets of global observables \cite{Reimann08,Short10}.

\headline{Maximum entropy principle}
%
In his seminal papers \cite{Jaynes57,Jaynes57_2} E.\ T.\ Jaynes suggested a \emph{maximum entropy principle} as a possible foundation for statistical mechanics. In short, Jaynes argues that the correct method to calculate expectation values of observables that give only limited knowledge about a system is to take the state with maximum entropy among the configurations that are compatible with our partial knowledge. His argument is based on information theoretic considerations of the method of statistical inference.
 
Surprisingly, the unitary dynamics of quantum mechanics does also naturally imply a maximum entropy principle that is quite similar in style. Recently it was shown in Ref.~\cite{absence} that whenever an expectation value of some observable equilibrates in the sense defined above, it equilibrates towards the expectation value it would have in the state that maximizes the von Neumann entropy among the states that have the same expectation values for all conserved quantities, i.e., of all observables that commute with the Hamiltonian. Again the result is derived without making any special assumptions and without approximations. In contrast to the common approach in which the maximum entropy principle appears as an axiom, this result follows directly from first principles when we chose a microscopic quantum description. What makes the maximum entropy principle from quantum dynamics different form the usual Jaynes' principle is that the number of conserved quantities of a quantum system grows with the dimension of the Hilbert space and thus exponentially with the system size. Contrarily, Jaynes formulated his maximum entropy principle having situations with partial knowledge about a handful of natural physical observables in mind. How exactly these two maximum entropy principles are related is yet to be explored in more detail.

\headline{Thermalization}
%
One of the most important applications of equilibrium statistical physics is to calculate the properties of systems at a well defined temperature.The standard assumption going into these calculations is that the state of such a system is described by a Gibbs state. The Gibbs state and the canonical ensemble can be derived from the equal a priory probability postulate under certain assumptions about the density of states of a bath with which the system can exchange energy. Alternatively it is possible to justify the Gibbs state by using Jaynes' maximum entropy principle, showing that the Gibbs state is the state that maximizes the conditional entropy given a fixed energy expectation value. However, it remains unclear how, and under which conditions, subsystems of quantum systems actually thermalize, by which, in this section, we mean that it equilibrates towards a Gibbs states with a well defined temperature. Note that it is not easy to relate the notion of thermalization we use in this section to the detailed balance condition used throughout the rest of this article. 

Earlier works attempting to solve this problem \cite{Srednicki,tasaki98,Goldstein06} either rely on certain unproven hypotheses such as the so-called \emph{eigenstate thermalization hypothesis}, or they are restricted to quite special situations such as coupling Hamiltonians of a special form, or they merely prove typicality arguments instead of dynamical relaxation towards a Gibbs state. Although the results obtained in these papers are very useful and have significantly improved our understanding of the process of thermalization, they do not yet draw a complete and coherent picture.

An attempt to settle the question of thermalization will be made in a forthcoming paper~\cite{Arnau11}. As discussed above we already know conditions under which we can rigorously guarantee equilibration \cite{Reimann08,Linden09,Short10}. What remains to be done is to identify a set of conditions under which one can guarantee that the equilibrium state of a subsystem is close to a Gibbs state. By using a novel perturbation theory argument and carefully bounding all the errors in an approximation similar to that of \cite{Goldstein06} one can indeed identify such a set of sufficient (and more or less necessary) conditions, that can be summarized in a non-technical way as follows:
\begin{itemize}
\item[(i)] The energy content and the Hilbert space dimension of the bath must be much larger than the respective quantities of the system.
\item[(ii)] The coupling between them must be strong enough, in particular much stronger than the gaps of the decoupled Hamiltonian. This ensures that the eigenbasis of the full Hamiltonian is sufficiently entangled (a lack of entanglement provably prevents thermalization \cite{absence}).
\item[(iii)] At the same time, the coupling must be weak in the sense that it is much smaller than the energy uncertainty of the initial state. This is a natural counterpart to the weak coupling assumption known from classical statistical mechanics.
item[(iv)] The energy uncertainty of the initial state must be small compared to the energy content and at the same time large compared to the level spacing.
Moreover, the energy distribution must satisfy certain technical smoothness conditions.
\item[(v)] The spectrum of the bath must be well approximable by an exponential on the scale of the energy uncertainty and the density of states must grow faster than exponential. This property of the bath is ultimately the reason for the exponential form of the Gibbs state and is also required in the classical derivation of the canonical ensemble. Most natural many particle systems have this property.
\end{itemize}
In summary, one can say that more or less the same conditions that are used in the classical derivation of the canonical ensemble appear naturally in the proof of dynamical thermalization.

\headline{Time scales}
%
The most important open problem for the approach described above is that rigorous bound on the time scales for decoherence/equilibration/thermalization are not yet known. The results derived in \cite{Reimann08,Linden09,Linden10,PRE81,absence,Short10} only tell us that decoherence/equilibration/thermalization must eventually happen under the given conditions, but they do not tell us how long it takes. In general this seems to be tough question, but for exactly solvable models the time scales can be derived \cite{Cramer}.

\section{Entropy production}
\label{sec:entropy}

Returning to the classical framework, let us now study the problem of entropy production. As outlined in Sect. 2, thermalizing systems (i.e. systems with balanced rates relaxing into thermal equilibrium) can contain subsystems which are out of thermal equilibrium in the sense that the transition rates $\wssp$ do not obey detailed balance. The apparent contradiction is resolved by observing that the effective rates in the subsystem are generally time-dependent and will eventually adjust in such a way that the subsystem thermalizes as well. However, for a limited time it is possible to keep them constant in such a way that they violate detailed balance. This is exactly what happens in experiments far from equilibrium -- typically they rely on external power and will quickly thermalize as soon as power is turned off.

\begin{figure}
\centering\includegraphics[width=70mm]{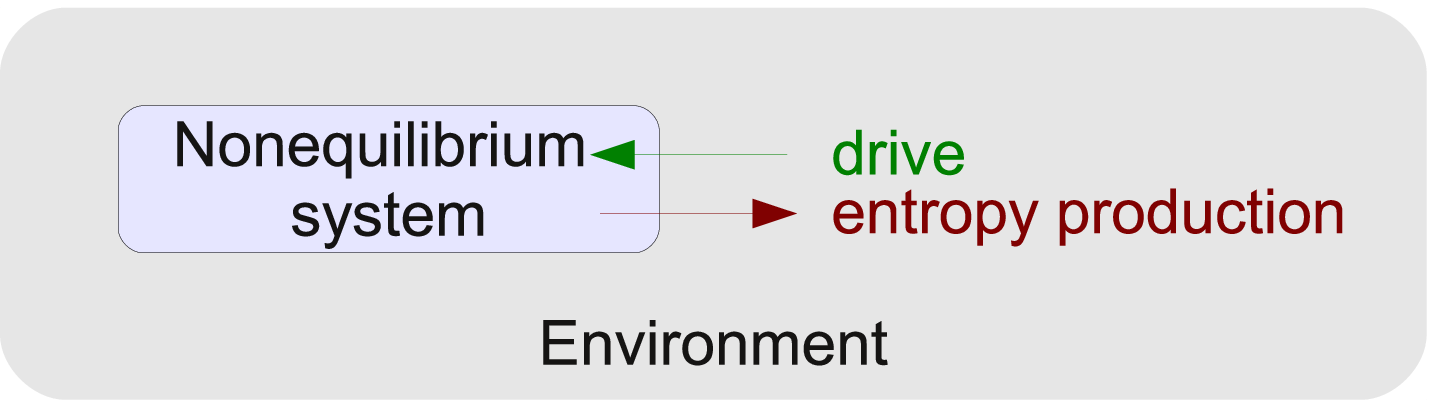}
\label{fig:entropyproduction}
\smallcaption{
Entropy production: A non-thermalizing system cannot exist on its own but must be driven from the outside. The external drive, that keeps the system away from thermal equilibrium, inevitably increases the entropy in the environment.}
\end{figure}

The external drive which is necessary to keep a subsystem away from thermal equilibrium will on average increase the entropy in the environment, as sketched in Fig.~\ref{fig:entropyproduction}. In the following we discuss various attempts to quantify this entropy production.

\headline{Entropy changes}
%
For a subsystem embedded in an environment we distinguish three types of configurational entropies, namely, the configurational entropy of the total system (`Universe'), the entropy of the subsystem (experiment) and the entropy in its environment:
\begin{eqnarray}
\St(t,\c) &=& - \ln \Pc(t)\,, \\
\Ss(t,\s) &=& - \ln \Ps(t)\,, \\
\Se(t,\c) &=&  \St(t,\c) - \Ss(t,\pi(\c))\,.
\end{eqnarray}
Averaging over many realizations the corresponding mean entropies are given by
\begin{eqnarray}
\ASt(t) &=& \langle\St(t,\c)\rangle_\c  \;=\; - \sum_{\c\in\Ot} \Pc(t) \ln \Pc(t)\,, \\
\ASs(t) &=& \langle\Ss(t,\s)\rangle_\s  \;=\; - \sum_{\s\in\Os} \Ps(t) \ln \Ps(t)\,, \\
\ASe(t) &=& \langle\Se(t,\c)\rangle_\s  \;=\;  \St(t) - \Ss(t)\,.
\end{eqnarray}
The time derivative of these averages is given by
\begin{eqnarray}
\frac{\d}{\d t} \ASt(t) &=& \sum_{\c,\cp\in\Ot} \Jccp(t) \ln \frac{\Pc(t)}{\Pcp(t)}\,,\\
\label{eq:systemsav}
\frac{\d}{\d t} \ASs(t) &=& \sum_{\s,\sp\in\Os} \Jssp(t) \ln \frac{\Ps(t)}{\Psp(t)}\,,
\end{eqnarray}
where we used the master equations (\ref{eq:smaster}) and (\ref{eq:cmaster}).

Let us now consider a temporal regime in which the rates of the subsystem can be considered as constant. In this case the subsystem, which is often small compared to the environment, may quickly reach a non-thermalized stationary state, which is often referred to as \textit{non-equilibrium steady state} (NESS) in the literature. In this case the average entropy of the subsystem will saturate whereas the total system thermalizes according to the Second Law, meaning that the average entropy in the environment increases. This environmental entropy production is the price Nature has to pay for keeping a subsystem away from thermal equilibrium. 

To be more specific, let us now assume that the total system follows a particular stochastic path $\Gamma: t\mapsto c(t)$ 
\begin{equation}
\Gamma:\;\c_0 \to \c_1 \to \c_2 \to\ldots \qquad\mbox{at times}\qquad t_0, t_1, t_2,\ldots
\end{equation}
Whenever $\pi(\c_i)\neq\pi(\c_{i+1})$ a transition in the total system implies a transition in the subsystem, as sketched in Fig.~\ref{fig:sectors}. Denoting the corresponding transition times by $t_{n_i}$, the projected stochastic path of the subsystem $\gamma=\pi[\Gamma]$ reads
\begin{equation}
\label{eq:configss}
\gamma:\;\s_0 \to \s_1 \to \s_2 \to\ldots \qquad\mbox{at times}\qquad t_{n_0}, t_{n_1}, t_{n_2},\ldots
\end{equation}
where $s_i=\pi(c_{n_i})$. Along their respective stochastic paths the configurational entropies of the total system and the subsystem are given by
\begin{eqnarray}
\St^\Gamma(t)&=&- \ln \P_{\c(t)}(t)\,,\\
\Ss^{\gamma}(t)&=&- \ln \P_{\s(t)}(t)\,.
\end{eqnarray}
How do these quantities change with time? Following Ref.~\cite{Seifert} the temporal evolution of the configurational entropy is made up of a continuous contribution caused by the deterministic evolution of the master equation and a discontinuous contribution occurring whenever the system hops to a different configuration. This means that the time derivative of the systems entropy is given by
\begin{equation}
\label{eq:systmsentropychange}
\frac{\d}{\d t}\Ss^{\gamma}(t) \;=\; -\frac{\dot P_{s(t)}(t)}{P_{s(t)}(t)} - \sum_{j} \delta(t-t_{n_j}) \ln\frac{P_{\s_{j}}(t)}{P_{\s_{j-1}}(t)}\,.
\end{equation}
Similarly, the total entropy of the `Universe' is expected to change as
\begin{equation}
\label{eq:totalentropychange}
\frac{\d}{\d t}\St^\Gamma(t) \;=\; -\frac{\dot P_{c(t)}(t)}{P_{c(t)}(t)} - \sum_{n} \delta(t-t_n)\ln\frac{P_{c_n}(t)}{P_{c_{n-1}}(t)}
\end{equation}
so that the environmental entropy production is given by their difference:
\begin{equation}
\label{eq:environmentalentropychange}
\frac{\d}{\d t}\Se^\Gamma(t) \;=\; \frac{\d}{\d t}\Ss^{\gamma}(t) -\frac{\d}{\d t}\St^{\Gamma}(t)\,.
\end{equation}
This formula is exact but useless from a practical point of view because the actual stochastic trajectory $\Gamma$ of the total system (the whole `Universe') is generally not known.

\headline{Effective environmental entropy production}
%
Based on previous work by Andrieux and Gaspard~\cite{Gaspard}, Seifert suggested a very compact formula for the effective entropy production in the environment caused by the embedded subsystem~\cite{Seifert}:
\begin{equation}
\boxed{
\label{eq:entropyproductionformula}
\frac{\d}{\d t}\Se^{\gamma}(t) \;=\; - \sum_{j} \delta(t-t_{n_j}) \ln\frac{\w_{\s_j\s_{j+1}}(t)}{\w_{\s_{j+1}\s_j}(t)}\,.}
\end{equation}
This formula tells us that each transition $\s\to\sp$ in the subsystem causes an instantaneous change of the environmental entropy by the log ratio of the forward rate $\wssp$ divided by the backward rate $\wsps$. Together with Eq.~(\ref{eq:systmsentropychange}) this formula would imply that the \textit{total} entropy changes according to
\begin{equation}
\label{eq:vgl}
\frac{\d}{\d t}\St^{\gamma}(t) \;=\; -\frac{\dot P_{s(t)}(t)}{P_{s(t)}(t)} - \sum_{j} \delta(t-t_{n_j}) \ln\frac{P_{\s_{j}}(t)\w_{\s_j\s_{j+1}}(t)}{P_{\s_{j-1}}(t)\w_{\s_{j+1}\s_j}(t)}\,.
\end{equation}
This expression differs significantly from the exact formula~(\ref{eq:totalentropychange}) so that it can be only meaningful in an effective sense under certain conditions or in a particular limit. 

Before discussing these underlying assumptions in detail, we like to note that Eq.~(\ref{eq:entropyproductionformula}) is indeed very elegant. It does not require any knowledge about the nature of the environment, instead it depends exclusively on the stochastic path of the subsystem and the corresponding transition rates. Moreover, this quantity can be computed very easily in numerical simulations: Whenever the program selects the move $\s\to\sp$, all what has to be done is to  increase the environmental entropy variable by $\ln(\wssp/\wsps)$\footnote{In order to avoid unnecessary floating point operations, it is useful to store all possible log ratios of the rates in an array.}. Note that the logarithmic ratio of the rates requires each transition to be reversible. 

To motivate formula~(\ref{eq:entropyproductionformula}) heuristically, Seifert argues the corresponding \textit{averages} of the entropy production reproduce a well-known result in the literature. More specifically, he shows that Eq.~(\ref{eq:entropyproductionformula}) averaged over many possible paths gives the expression.
\begin{equation}
\frac{\d}{\d t} \ASe(t) \;=\; \sum_{\s,\sp\in\Os} \Jssp(t) \ln \frac{\wssp(t)}{\wsps(t)}\,.
\end{equation}
Combined with Eq.~(\ref{eq:systemsav}) one obtains the \textit{average} entropy production in the total system
\begin{equation}
\label{eq:avtot}
\frac{\d}{\d t} \ASt(t) \;=\; \sum_{\s,\sp\in\Os} \Jssp(t) \ln \frac{\Ps(t)\wssp(t)}{\Psp(t)\wsps(t)}\,.
\end{equation}
This formula was first introduced by Schnakenberg~\cite{Schnakenberg} and has been frequently used in chemistry and physics~\cite{Li}. It is in fact very interesting to see how Schnakenberg derived this formula. As described in detail in Appendix~A, he considered a fictitious chemical system of homogenized interacting substances which resemble the dynamics of the master equation in terms of particle concentrations. Applying standard methods of thermodynamics, he was able to prove Eq.~(\ref{eq:avtot}). The rational behind this derivation is to assume that the environment is always close to thermal equilibrium.

\headline{Limit of fast thermalization in the environment}
In the following we show that the entropy production formula is correct in the limit where the environment equilibrates immediately whenever a transition occurs in the subsystem. This requires a separation of time scales of the internal dynamics of the subsystem on the one hand and the relaxation in the environment on the other. 

As shown in Sect.~\ref{sec:setup}, the projection $\pi:~\Ot\to\Os$ divides the configuration space of the total system into sectors of configurations $c$ mapped onto the same $s$ (see Fig.~\ref{fig:sectors}). It is important to note that even if the dynamics of the total system was ergodic, the dynamics \textit{within} these sectors is generally non-ergodic, meaning that they may split up into various ergodic subsectors. For example, if the subsystem starting in a certain configuration returns to the same configuration after some time, the corresponding subsector may have changed, reflecting the change of entropy in the environment. For a given stochastic path of the small system, we will denote the corresponding subsector as $\Ot^{s(t)} \subset \Ot$.

Let us now assume that the environmental degrees of freedom thermalize almost instantan\-eously, reaching maximal entropy within the actual subsector. This means that the system quickly reaches a state where the probabilities
\begin{equation}
\label{eq:immediate}
\Pc(t)=\left\{
\begin{array}{cc}
\Ps(t)/N_s(t) & \mbox{ if } c \in \Ot^{s(t)}\\
0 & \mbox{ otherwise. }
\end{array}
\right.
\end{equation}
are constant on $\Ot^{s(t)}$. This implies that in the formula~(\ref{eq:environmentalentropychange}) for the environmental entropy production, namely
\begin{equation}
\frac{\d}{\d t} \Se^\Gamma(t) \;=\; \frac{\dot P_{c(t)}(t)}{P_{c(t)}(t)} -\frac{\dot P_{s(t)}(t)}{P_{s(t)}(t)} +\sum_{j} \delta(t-t_{n_j}) \ln\frac{P_{\s_{j}}(t)}{P_{\s_{j-1}}(t)}- \sum_{n} \delta(t-t_n)\ln\frac{P_{c_n}(t)}{P_{c_{n-1}}(t)}\,,
\end{equation}
the first two terms cancel. Therefore, this expression reduces to
\begin{equation}
\frac{\d}{\d t} \Se^\Gamma(t) \;=\; \sum_{j} \delta(t-t_{n_j}) \ln\frac{P_{\s_{j}}(t)}{P_{\s_{j-1}}(t)}- \sum_{n} \delta(t-t_n)\ln\frac{P_{\pi(c_n)}(t)N_{\pi(c_{n-1})}(t)}{P_{\pi(c_{n-1})}(t)N_{\pi(c_n)}(t)}\,.
\end{equation}
Obviously, only those terms in the second sum will contribute where $\pi(c_{n-1})\neq \pi(c_n)$, i.e. where $n=n_j$, hence the sum can be reorganized as
\begin{equation}
\label{eq:calculation}
\frac{\d}{\d t} \Se^\gamma(t) \;=\; \sum_{j} \delta(t-t_{n_j})\left[\frac{P_{\s_{j}}(t)}{P_{\s_{j-1}}(t)}-\ln\frac{P_{s_j}(t)N_{s_{j-1}}(t)}{P_{s_{j-1}}(t)N_{s_j}(t)} \right]
\;=\; \sum_{j} \delta(t-t_{n_j})\ln\frac{N_{s_j}(t)}{N_{s_{j-1}}(t)}\,
\end{equation}
which now depends only on the stochastic path $\gamma$ of the subsystem. This result, saying that the entropy increase is given by the logarithmic ratio of the number of available configurations, is very plausible under the assumption of instantaneous thermalization of the environment. It remains to be shown that this ratio is related to the ratio of the effective rates. In fact, inserting (\ref{eq:immediate}) into (\ref{eq:reducedrate}) we obtain
\begin{equation}
\wssp(t) \;=\;  \frac{\sum_{\c(\s)} \Pc(t)\sum_{\cp(\sp)}\wccp}{\sum_{\c(\s)}\Pc(t)}
=  \frac{\sum_{\c(\s)} \sum_{\cp(\sp)}\wccp}{N_s(t)}\,,
\end{equation}
hence $\wssp/\wsps = N_{s'}/N_s$. Inserting this relationship into Eq.~(\ref{eq:calculation}) we arrive at the formula for the effective entropy production~(\ref{eq:entropyproductionformula}). This proves that this formula is valid under the assumption that the environmental degrees of freedom thermalize instantaneously after each transition of the subsystem.

\headline{Conjecture for systems in a non-thermalized environment}
What happens if the environment does not thermalize immediately after each transition of the subsystem? To answer this question we performed numerical simulations of a three-state system in an environment with 50 configurations and randomly chosen transition rates. The results (not shown here) suggest that the entropy production predicted by formula~(\ref{eq:entropyproductionformula}) deviates from the true entropy production in both directions. However, averaging over many independent stochastic paths while keeping the rates fixed, the formula seems to systematically underestimate the actual entropy production. This leads us to the conjecture that formula~(\ref{eq:entropyproductionformula}) may serve as a lower bound for the expectation value of the entropy production. 

\section{Fluctuation theorem revisited}
\label{sec:fluc}

The Second Law of thermodynamics tells us that the entropy $\St$ of an isolated system increases on average during thermalization, approaching the maximal value $\St=\ln|\Ot|$. The fluctuation theorem generalizes this statement by studying the actual fluctuations of the total entropy, including the Second Law as a special case~\cite{evans93,evans94,gallavotti95,kurchan98,lebowitz99,maes99,Jia1,harris07,kurchan07}. In this section we want to suggest an alternative particularly transparent proof of the fluctuation theorem.

\begin{figure}
\centering\includegraphics[width=160mm]{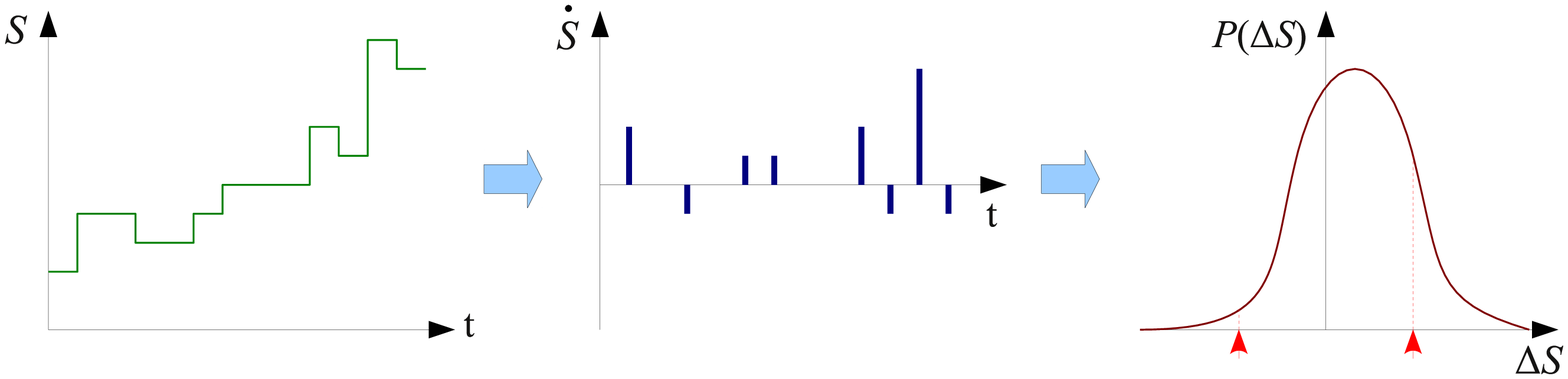}
\vspace{-10mm}
\label{fig:fluctheo}
\smallcaption{ 
Configurational entropy of an isolated system, its temporal derivative and the corresponding probability distribution of entropy differences.
}
\end{figure}

The generic situation is sketched in Fig.~\ref{fig:fluctheo}. Although the average entropy of an isolated system will increase, the actual configurational entropy will not grow monotonously, instead it jumps discontinuously by finite differences $\Delta\St$. As the average entropy increases, these differences will be preferentially positive, but sometimes fluctuations in opposite direction may occur. Depending on the system under consideration, these differences will be distributed according to a certain asymmetric distribution. The fluctuation theorem states that this distribution is constrained by the condition
\begin{equation}
\frac{P(\Delta\St)}{P(-\Delta\St)}=e^{\Delta\St}\,.
\end{equation}
It relates pairs of values at opposite locations on the abscissa, as marked by the red arrows in Fig.~\ref{fig:fluctheo}. In other words, given one half of the distribution, the fluctuation theorem predicts the other half.

The fluctuation theorem holds for any stochastic system and is usually proved by comparing the entropy production along a given stochastic path with the entropy production along the reverse path. Here we suggest an alternative simple proof. Starting point is the observation that the fluctuation relation is invariant under convolution, i.e. if two function $f,g$ satisfy the property $f(x)=e^xf(-x)$ and $g(x)=e^xg(-x)$ their convolution product will satisfy the same property:
\begin{eqnarray}
(f*g)(x) &=& \int \d y\, f(y)g(x-y) \;=\; \int \d y\, e^yf(-y)e^{x-y}g(-x+y) \\
&=& e^x \int \d y\, f(-y)g(y-x) \;=\; e^x \int \d y\,f(y)g(-y+x) \;=\; e^x (f*g)(-x)\,. \nonumber
\end{eqnarray}
This means that the sum of random variables obeying the fluctuation relation will again obey the fluctuation relation. The remaining proof consists of two steps:

\begin{enumerate}
    \item First we prove the fluctuation theorem for a \textit{single} transition. According to Eq.~(\ref{eq:totalentropychange}), an \textit{individual} transition $\c\to\cp$ changes the configurational entropy of an isolated system by $\Delta\St=\ln\Jccp-\ln\Jcpc$. Since transition occurs with frequency $\Pc\wccp=\Jccp$, we have $P(\Delta\St)=\Jccp$ and similarly $P(-\Delta\St)=\Jcpc$ for the reverse transition. Therefore the fluctuation relation
    \begin{equation}
    \Delta\St=\ln\Bigl(P(\Delta\St)/P(-\Delta\St)\Bigr)\,
    \end{equation}
    holds trivially for a single transition.\vspace{1mm}
    \item  The entropy change $\Delta S$ over a finite time is the sum of entropy changes caused by individual transition. Summing random variables means to convolve their probability distributions. Since the fluctuation theorem is invariant under convolution, it follows that this sum will automatically obey the fluctuation relation as well.
\end{enumerate}
As well-known, the fluctuation theorem implies the nonequilibrium partition identity
\begin{equation}
\langle\exp(-\Delta \St)\rangle = 1
\end{equation}
as well as the Second Law of thermodynamics
\begin{equation}
\langle\Delta \St\rangle \geq 0.  
\end{equation}
Note that the fluctuation theorem holds exactly only in isolated systems. However, it may also hold approximately in the limit $t\to\infty$ for the environmental entropy under certain assumptions if the subsystem is stationary.

\section{Concluding remarks}
\label{sec:conclusions}

In this paper we have addressed several aspects of classical non-equilibrium statistical physics, describing its general setup and its justification from the quantum perspective. In particular, we have focused on the problem of entropy production. As we have pointed out, the commonly accepted formula for entropy production in the environment $\Delta\Se = \ln(\wssp/\wsps)$ holds only in situations where the environment thermalizes almost immediately after each transition of the subsystem. Whether this separation of time scales is valid in realistic situations remains to be seen. Moreover, we have suggested a conjecture that this formula gives a lower bound to the average entropy production in the environment.

\appendix
\section{Tracing the historical route to entropy production}
\label{sec:AppA}

It is instructive to retrace how the formula for entropy production was derived by Schnakenberg in 1976~\cite{Schnakenberg}. To quantify entropy production, Schnakenberg considers a fictitious chemical system that mimics the dynamics of the master equation. This fictitious system is based on the following assumptions:
\begin{itemize}
  \item Each configuration $\c$ of the original system is associated with a chemical species $X_\c$ in an ideal homogeneous mixture of molecules.
  \item The molecules react under isothermal and isochoric conditions by $X_\c\rightleftharpoons X_\cp$ in such a way that their concentrations $\C=N_\c/V$ evolve in the same way as the probabilities in the master equation, i.e.
    \begin{equation}
    \frac{\d}{\d t} \C \;=\; \sum_{\cp} \Bigl(\Cp\wcpc-\C\wccp\Bigr)\,.
    \end{equation}
  \item The reactions are so slow that standard methods of thermodynamics can be applied.
\end{itemize}
Under isothermal and isochoric conditions the chemical reactions change the particle numbers $N_i$ in such a way that the Helmholtz free energy $F$ is maximized. In chemistry the corresponding thermodynamic current is called the \textit{extent of reaction} $\extent$, which is defined as the expectation value of the accumulated number of forward reactions $X_\c \rightarrow X_\cp$
minus the number of backward reactions $X_\c \leftarrow X_\cp$. Note that $\extent$ does not account for fluctuations, instead it is understood as a macroscopic deterministic quantity that grows continuously as
\begin{equation}
\dot\extent \;=\; \Nc \wccp - \Ncp \wcpc\,.
\end{equation}
According to conventional thermodynamics, a thermodynamic flux is caused by a conjugate thermodynamic force which is the partial derivative of the thermodynamic potential with respect to the flux. In chemistry the thermodynamic force conjugate to the extent of reaction  $\dot\extent$ is the so-called \textit{chemical affinity}
\begin{equation}
\affin \;=\; \left.\frac{\partial F}{\partial \extent}\right|_{V,T} \,.
\end{equation}
With this definition the temporal change of the free energy is given by
\begin{equation}
\label{eq:Fchange}
\dot F \;=\; \sum_{\c\cp} \affin \dot\extent\,.
\end{equation}
The affinity is related to the chemical potential of the involved substances as follows. On the one hand, the reaction changes the particle number by $\dot\Nc=-\dot\extent$ and $\dot\Ncp=+\dot\extent$. On the other hand, the change of the free energy can be expressed as $\dot F=\sum_\c\frac{\partial F}{\partial \Nc} \dot \Nc=\sum_\c\mu_\c\dot\Nc$. Comparing this expression with Eq.~(\ref{eq:Fchange}) the affinity can be expressed as
\begin{equation}
\affin = \mu_\cp-\mu_\c \,.
\end{equation}
For an ideal mixture the Helmholtz free energy is given by 
\begin{equation}
F \;=\; \sum_\c \Nc (q_\c + k_BT\ln \Nc)\,,
\end{equation}
where $q_c$ is a temperature-dependent constant. The chemical potential of species $X_c$ is
\begin{equation}
\mu_\c \;= \; \frac{\partial F}{\partial \Nc} \;=\; \mu_\c^0 + k_BT \ln N_\c
\end{equation}
with $\mu_\c^0=q_\c+k_BT$, so that the affinity is given by
\begin{equation}
\affin = \mu_\c^0 -\mu_\cp^0 + k_BT \ln \frac{\Ncp}{\Nc}\,.
\end{equation}
The fictitious chemical system relaxes towards an equilibrium state that corresponds to the stationary state of the original master equation. In this state the particle numbers $\Nc$ attain certain stationary equilibrium values $\Nc^{\rm eq}$. Moreover, the thermodynamic flux and its conjugate force vanish in equilibrium:
\begin{equation}
\affin^{\rm eq} = \extent^{\rm eq} = 0 \,.
\end{equation}
Because of $\affin^{\rm eq} =0$ we have
\begin{equation}
\mu_\c^0 -\mu_\cp^0 = k_BT \ln \frac{\Ncp^{\rm eq}}{\Nc^{\rm eq}}\,,
\end{equation}
which allows one to express the affinity as
\begin{equation}
\label{eq:Nquotient}
\affin = k_B T \ln \frac{\Ncp\Nc^{\rm eq}}{\Nc\Ncp^{\rm eq}}\,.
\end{equation}
On the other hand, $\extent^{\rm eq} = 0$ implies that
\begin{equation}
\frac{\Nc^{\rm eq}}{\Ncp^{\rm eq}} \;=\; \frac\wcpc\wccp\,.
\end{equation}
Inserting this relation into Eqs.~(\ref{eq:Nquotient}) and~(\ref{eq:Fchange}) the change of the free energy (caused by all reactions $X_\c\rightleftharpoons X_\cp$) is given by
\begin{equation}
\dot F \;=\; k_B T \sum_{\c\cp} \extent \ln \frac{\Ncp\wcpc}{\Nc\wccp}\,.
\end{equation}
Since temperature $T$ and internal energy $U$ of the mixture remain constant the variation of the free energy $F=U-TS$ is fully absorbed in a change of the entropy, i.e. $\dot F=-T\dot S$. This allows one to derive a formula for the entropy production
\begin{equation}
\label{eq:mainresult}
\dot S = - k_B \sum_{\c\cp} \extent \ln \frac{\C\wccp}{\Cp\wcpc}\,,
\end{equation}
where we used the fact that $\Nc/\Ncp=\C/\Cp$.

Since the extent of reaction $\extent$ just counts the number of reactions from $\c$ to $\cp$, this formula is just the continuum limit of Eq.~(\ref{eq:vgl}), proving the formula for entropy production~(\ref{eq:entropyproductionformula}). However, it is important to recall the underlying assumptions. Each component of the fictitious chemical systems was assumed to be internally equilibrated and methods of ordinary thermodynamics were used to derive Eq.~(\ref{eq:mainresult}). This means that the formula will only be meaningful in the limit where the environment equilibrates almost immediately. 

\newpage
\ack

This work was initiated at conference STATPHYS-VII in Kolkata, India. HH would like to thank the organizers for the warm hospitality and for the lively and productive atmosphere. CG and PJ thank the Centre for Quantum Technologies, Singapore, where part of this work was done.

\section*{References}

\providecommand{\newblock}{}

\end{document}